\documentclass[pra,amsmath,amssymb,twocolumn,preprintnumbers,showpacs]{revtex4-1}
\usepackage{graphicx}
\usepackage{epsfig}
\usepackage{bm}
\usepackage{tabularx}

\newcommand{\beq}{\begin{equation}}
\newcommand{\eeq}{\end{equation}}
\newcommand{\bea}{\begin{eqnarray}}
\newcommand{\eea}{\end{eqnarray}}
\newcommand{\Tr}{\text{Tr}}

\begin{document}
\preprint{LA-UR-12-20047}

\title{Improved lattice operators for non-relativistic fermions}
\author{Joaqu\'{\i}n E. Drut}
\affiliation{Theoretical Division, Los Alamos National Laboratory, Los Alamos, NM 87545-0001, USA}

\date {\today}

\begin{abstract}
In this work I apply a recently proposed improvement procedure, originally conceived to reduce finite lattice 
spacing effects in transfer matrices for dilute Fermi systems, to tuning operators for the calculation of 
observables. I construct, in particular, highly improved representations for the energy and the contact, as a 
first step in an improvement program for finite-temperature calculations. I illustrate the effects of improvement 
on those quantities with a ground-state lattice calculation at unitarity.
\end{abstract}

\pacs{03.75.Ss, 05.10.Ln, 12.38.Gc, 71.10.Fd}

\maketitle

%
\section{Introduction}

One of the main sources of uncertainty in Monte Carlo (MC) calculations of lattice field theories stems from 
finite lattice spacing effects~\cite{Latticebooks}. This is true in particular of non-relativistic systems at 
finite density, where 
the continuum limit is approached by taking the limit of diluteness, such that the interparticle distance
$\sim k_F^{-1}$ (where $k^{}_F$ is the Fermi momentum) is much larger than the lattice spacing $l$.
This is achieved in practice by first considering large 
volumes at fixed particle density, and then taking the zero-density limit. While this process is well defined, it 
is cumbersome to carry out, in part because calculations in large volumes require substantial amounts of
CPU time. 
It is therefore useful to consider alternative approaches, based on effective field theory and renormalization 
group concepts~\cite{Lepage, ImprovementQCD}, which treat lattice-spacing effects by modifying the
ultraviolet (UV) dynamics of the theory. In such formulations, lattice-spacing effects are eliminated at finite volume, 
even for densities that are not low by conventional standards.

In recent work, Endres et al. \cite{Endresetal, Endresetal2} proposed a novel way to systematically reduce 
lattice-spacing effects in calculations of non-relativistic fermions. The method enables tuning of the 
lattice theory to high accuracy, such that the low end of the continuum energy 
spectrum is reproduced for the desired values of the two-body scattering parameters in the effective range expansion:
\beq
\label{EffectiveRangeExpansion}
p \cot \delta(p) = -\frac{1}{a}  + \frac{1}{2} r^{}_\text{eff} p^2 + O(p^4),
\eeq
where $\delta$ is the scattering phase shift, $a$ is the scattering length, and $r^{}_\text{eff}$ is the effective range.
The connection between the bare lattice theory (or rather its two-body spectrum) and these physical parameters 
is given by L\"uscher's formula~\cite{Luescher}, which relates the phase shift to the energy $E = p^2/m$ of 
the two-body problem in a box of side $L$:
\beq
\label{Luscherformula}
p \cot \delta(p) = \frac{1}{\pi L}\mathcal S^{}(\eta)
\eeq
where $\eta = \frac{pL}{2\pi}$ and
\beq
\label{Seta}
\mathcal S^{}(\eta) \equiv \lim_{\Lambda \to \infty} \left ( \sum_{\bf n}^{} 
\frac{\Theta(\Lambda^2 - {\bf n}^2)}{{\bf n}^2 - \eta^2} - 4 \pi \Lambda \right ),
\eeq
where the sum is over all 3D integer vectors (the evaluation of $\mathcal S^{}(\eta)$ is discussed in Appendix A),
and $\Theta(x)$ is the Heaviside function. Throughout this work, we shall take units such that $\hbar = k_B^{} = m = 1$, 
where $m$ is the mass of the fermions.

The two-body matching condition described above completely specifies the physics of dilute systems, such 
as those currently realized with ultracold atomic gases (see e.g. Ref.~\cite{Experiments} for a review of the 
experimental situation). In this work we shall restrict ourselves to such systems, neglecting the additional
complications that arise for instance at higher densities or for nuclear systems, where three- and higher-body
forces play an important role.

We shall briefly review the work of Ref.~\cite{Endresetal} below, but it is useful to mention the main steps 
underlying the method at this point. First, one writes down the transfer matrix 
$\mathcal T$, representing the two-body interaction via a generalized Hubbard-Stratonovich (HS) 
transformation~\cite{HS, NegeleOrlandbook}. 
The latter contains $N^{}_{\mathcal O}$ arbitrary coefficients $C_n^{}$. The matrix elements of $\mathcal T$ 
are then computed in the 
subspace of two particles at vanishing total momentum. The resulting matrix is then diagonalized in the 
$s$-wave subspace (or rather its lattice equivalent), and the HS coefficients $C_n^{}$ are tuned using an iterative 
algorithm such that the eigenvalues of the proposed $\mathcal T$ match $\exp(-\tau E)$, where $\tau$ is the
temporal discretization parameter and $E$ are the energies dictated by L\"uscher's formula, for the desired box size $L$ 
and choice of scattering parameters.

In the case of the unitary limit, where by definition 
\beq
p \cot \delta(p) \equiv 0, 
\eeq
the above procedure yields
considerable improvement in the approach to the continuum. Indeed, with a single coefficient $C^{}_0$ 
(the simplest case, discussed in Sec.~\ref{SimpleExample}) one may tune the scattering length $a$, but the 
effective range $r^{}_\text{eff}$ remains finite due to lattice-spacing artifacts. 
Naturally, by introducing more parameters one may tune the effective-range expansion in 
Eq.~(\ref{EffectiveRangeExpansion}) to higher accuracy, thus systematically eliminating the need for extrapolations to the 
dilute limit and leaving only finite-volume effects unaccounted for. 

It is the main objective of this work to extend the approach of Ref.~\cite{Endresetal} to designing not only improved 
transfer matrices but also improved operators for the calculation of observables. This is an important step toward 
reducing lattice-spacing effects in finite-temperature calculations. While these effects were studied at unitarity in 
Ref.~\cite{Priviteraetal} for the Hubbard model using dynamical mean-field theory, careful systematic studies in 
full-fledged MC calculations have only recently started to appear~\cite{Endresetal, 
Endresetal2, Carlsonetal, Leeetal} and are still restricted to the ground state. Interestingly, recent studies based
on functional renormalization group techniques \cite{Braun} have also started to tackle the problem of extrapolating
to infinite volume and particle number in the unitary limit.

In Sec.~\ref{SimpleExample} we analyze a basic example, as a primer to reviewing the more 
sophisticated formalism of Ref.~\cite{Endresetal}, which we explain briefly in Sec.~\ref{ImprovedT}.
In Sec.~\ref{ImprovedO} we present the main developments of this work, with the corresponding illustrative
results and conclusions appearing in Sec.~\ref{Results}. Finally, some of the less trivial numerical issues are 
explained in the Appendix.

\section{\label{SimpleExample}A simple example}

Before proceeding to a more detailed discussion, let us analyze a simple case, to fix notation as well as ideas.
Consider the following lattice Hamiltonian:
\beq
\hat H \equiv \!\!\!
\sum_{\bm{k},s=\uparrow,\downarrow} \!\!
\frac{k^2}{2m} \,
\hat a^{\dagger}_{s,\bm{k}} \, \hat a^{}_{s,\bm{k}}
-\,g \sum_{\bm{i}} \hat n^{}_{\uparrow,\bm{i}} \, \hat n^{}_{\downarrow,\bm{i}},
\eeq
where $s$ denotes the spin projection, $g$ is the 
bare lattice coupling constant, and
$\hat{n}_{s,\bm{i}}^{} \equiv\hat{a}_{s,\bm{i}}^\dagger\, \hat{a}_{s,\bm{i}}^{}$ 
denotes the number density operator at lattice position $\bm{i}$ for spin projection~$s$.
The value of $g$ is tuned to the desired two-body scattering properties by solving the 
two-body problem on the lattice and finding the scattering amplitude, which in this case can 
be done analytically. 

The transfer matrix is then expressed approximately in powers of the
imaginary time step $\tau$ using a Suzuki-Trotter decomposition, for instance as follows:
\beq
\label{Tmatrix}
\mathcal T \equiv e^{-\tau \hat H} \simeq 
e^{-\frac{\tau \hat T}{2}}
e^{-\tau \hat V}
e^{-\frac{\tau \hat T}{2}} + O(\tau^2),
\eeq
where
\bea
\hat T \equiv \sum_{\bm{k}} && \!\!\!\!
{\hat T}^{}_\uparrow + {\hat T}^{}_\downarrow, \quad 
{\hat T}^{}_s \equiv \sum_{\bm{k}} \frac{k^2}{2m}
\hat a^{\dagger}_{s,\bm{k}} \, \hat a^{}_{s,\bm{k}}, \\
&&\!\!\!\!\!\!\!\!\!\!\!\! \text{and} \quad
\hat V \equiv -\,g \sum_{\bm{i}} \hat n^{}_{\uparrow,\bm{i}} \, \hat n^{}_{\downarrow,\bm{i}}.
\eea
The kinetic energy factor in Eq.~(\ref{Tmatrix}) is easy to treat, as it is an exponential of a 
one-body operator which is diagonal in momentum space. Notice that we have chosen to
define the dispersion relation $E = k^2/2m$ in momentum space, as opposed to defining it 
via a discrete representation of the Laplacian operator in coordinate space (which is common
in Hubbard-model type formulations). 

On the other hand, the potential energy factor represents a challenge,
as it is the exponential of a non-trivial two-body operator (as the original Hamiltonian). 
It is at this point that the HS transformation
plays a central role, allowing us to exchange the complexity of the two-body operator for a path integral 
involving only one-body operators. Specifically, we write
\bea
\label{HSTLee}
\exp(-\tau \hat V) &&\!\!\!\!\!\!= \prod_{\bm{i}} \exp(\tau g  \hat n^{}_{\uparrow,\bm{i}} \, \hat 
n^{}_{\downarrow,\bm{i}})
\\ \nonumber 
&&\!\!\!\!\!\!\!\!\!\!\!\!\!\!\!\!\!\!\!\!= \!\!\int \!\!\mathcal D \sigma 
\prod_{\bm{i}} \left(1 + \sqrt{A}\; \hat n^{}_{\uparrow,\bm{i}} \sin \sigma_{\bm{i}}\right)
\left(1 + \sqrt{A}\; \hat n^{}_{\downarrow,\bm{i}} \sin \sigma_{\bm{i}}\right)
\eea
where $A = 2(e^{\tau g} - 1)$, $\mathcal D \sigma = \prod_{\bm{i}} d\sigma_{\bm{i}}/(2\pi)$, and 
$\sigma_{\bm{i}}$ is an external auxiliary field. In this way, one decouples the transfer matrix for each spin, 
such that we may write
\beq
\label{T_HS1}
\mathcal T = \int \mathcal D \sigma\ \mathcal T^{}_{\uparrow}[ \sigma] \mathcal T^{}_{\downarrow}[ \sigma]
\eeq
where, up to order $\tau^2$,
\beq
\label{T_HS2}
\mathcal T^{}_{s}[ \sigma] = 
e^{-\frac{\tau \hat T^{}_{s}}{2}}
\prod_{\bm{i}}\left(1 + \sqrt{A}\; \hat n^{}_{s,\bm{i}} \sin \sigma_{\bm{i}} \right)
e^{-\frac{\tau \hat T^{}_{s}}{2}},
\eeq
which in the one-particle subspace reduces to
\beq
\label{T_HS2sp}
\mathcal T^{}_{s}[ \sigma] = 
e^{-\frac{\tau \hat T^{}_{s}}{2}}
\left(1 + \sqrt{A}\; \sum_{\bm{i}} \hat n^{}_{s,\bm{i}} \sin \sigma_{\bm{i}} \right)
e^{-\frac{\tau \hat T^{}_{s}}{2}}.
\eeq
Notice that in Eq.~(\ref{HSTLee}) we have used a version of the HS transform due to Lee~\cite{DeanLee}, 
in which the auxiliary field is continuous and compact (the integral is restricted to the interval $[-\pi,\pi]$), 
as opposed to more conventional versions that are discrete, or continuous but unbounded.
Apart from the path integral, at this point applying the transfer matrix becomes a problem of applying a product of
one-body operators, which one can easily deal with.

All of the above steps are standard in the literature of many-body MC calculations, except perhaps for the use 
of a ``perfect" dispersion relation defined in momentum space, which has become more common only in recent 
years~\cite{BDM, Carlsonetal}. This feature can be regarded as a basic kind of improvement, as it reduces 
lattice-spacing effects relative to Hubbard-model approaches, where $dE/dk \to 0$ at high momenta.

This example captures the main features of most modern many-fermion MC calculations.
Nevertheless, the simplicity of this approach is in some ways excessive, largely because we only have
one coupling constant at our disposal. Indeed, should we desire to fix more than one coefficient in the 
effective-range expansion, we would need a richer bare interaction, and a correspondingly more sophisticated 
HS transformation. In this simple case, to reach for instance the unitary limit, we can tune the scattering length, 
but we also need to consider very dilute systems to avoid the effects of finite range induced by the 
UV lattice cutoff $\pi/l$, as in that case one has $r^{}_\text{eff} \simeq 0.40l$. Moreover, taking for example 
80 particles in a $10^3$ lattice volume, which shall be our illustrative example below, one finds $k_F^{} r^{}_\text{eff} \simeq 0.54$, 
which is not nearly as small as one would like.

As mentioned before, one can resort to alternative approaches that modify the Hamiltonian by including 
higher-order terms in a low-momentum expansion, tuning the corresponding couplings to eliminate UV effects. 
The latter strategy is essentially what Ref.~\cite{Endresetal} advocates, following the spirit of the Lattice QCD 
program initiated by Symanzik many years ago~\cite{ImprovementQCD} to design improved effective actions 
that approach the continuum limit (see e.g. Ref.~\cite{MILC_RMP}). Since it will be useful for the rest of this work, 
we shall briefly review the method of Ref.~\cite{Endresetal} in the next section.


\section{\label{ImprovedT}Improved transfer matrices}

The work of Ref.~\cite{Endresetal} tackles the problem of reducing UV lattice effects 
by defining an improved transfer matrix, using a
generalized HS transformation and a low-momentum expansion.
This is accomplished by allowing the constant $A$ to have a non-trivial
momentum dependence. For simplicity, it is useful to take $A({\bf p})$ to be diagonal 
in momentum space. Following Ref.~\cite{Endresetal}, we expand $A({\bf p})$ using a set of operators as
\beq
\label{Aexpansion}
A({\bf p}) = \sum_{n=0}^{N^{}_{\mathcal O}-1} C_{n}^{} {\mathcal O}_{n}^{}({\bf p}), 
\eeq
where one may choose to define, for convenience,
\beq
\label{Op1}
{\mathcal O}_{n}^{}({\bf p}) = \left(1 - e^{-{\bf p}^2}\right)^n,
\eeq
as done in Ref.~\cite{Endresetal} (up to an $n$-dependent constant in front), or, as we shall use in the rest of this work, 
\beq
\label{Op2}
{\mathcal O}_{n}^{}({\bf p}) = \left[2\sin({p}/{2})\right]^{2n}.
\eeq
Notice that this last expression contains only even powers of $p = \sqrt{{\bf p}^2}$, and is therefore
an analytic function of the momentum $\bf p$. Both of the above choices for ${\mathcal O}_{n}^{}({\bf p})$ behave
as $\sim p^{2n}$ at low momenta. For the purposes of Monte Carlo calculations, the operator $A({\bf p})$ can be 
computed once and for all at the beginning of the calculation, and is applied at each 
time slice using Fourier transforms.

In order to determine the coefficients $C_{n}^{}$, we find the explicit form of $\mathcal T$ for two particles, 
which upon integrating the auxiliary field is given in momentum space by
\bea
\label{2particleTmatrix}
\mathcal T_2^{}({\bm p}_\uparrow^{} {\bm p}_\downarrow^{};{\bm q}_\uparrow^{}{\bm q}_\downarrow^{}) 
&& \!\!\!\!\!\! = e^{-\frac{\tau T(p)}{2}}\! \left [
\delta_{{\bm p}_\uparrow^{} {\bm q}_\uparrow^{}}\delta_{{\bm p}_\downarrow^{} {\bm q}_\downarrow^{}} \right.
\nonumber \\
&&\!\!\!\!\!\!\!\!\!\!\!\!\!\!\!\!\!\!\!\!\!\!\!\!\!\!\!\!\!\!\!\!\!\!\!\!\!
\left . + 
\frac{\sqrt{A({\bm p}_\uparrow^{})} 
\sqrt{A({\bm p}_\downarrow^{})}} 
{2V}
\delta_{{\bm p}_\uparrow^{} + {\bm p}_\downarrow^{}, {\bm q}_\uparrow^{} + {\bm q}_\downarrow^{}}\right]
e^{-\frac{\tau T(q)}{2}}\!\!,
\eea
where $V$ is the lattice volume and 
\mbox{$T(k) = ({\bm k}_\uparrow^{2} + {\bm k}_\downarrow^{2})/2m$}.
One cannot fail to notice that the above transfer matrix is not Galilean invariant. We elaborate on this issue
in the appendices. In this work we shall take $\Lambda = \pi (1 - 10^{-5})$, following the steps of Ref.~\cite{Endresetal}. 
The operators in Eqs.~(\ref{Op1}) and~(\ref{Op2}) are defined to be constant above $p^2 = \Lambda^2$,
and the kinetic energy factor $\exp(-\tau T/2)$ is defined to vanish above that boundary, such that there
is no propagation for $p^2 \ge \Lambda^2$.

Evaluating $\mathcal T_2^{}$ in the center-of-mass frame, we have
\bea
\label{2particleTmatrixCOM}
\mathcal T_2^{}({\bm p}_r^{};{\bm q}_r^{}) 
= e^{-\frac{\tau p^{2}_r}{2m}} 
\left [
\delta_{{\bm p}_r^{} {\bm q}_r^{}}\! + \frac{A({\bm p}_r^{})}{2V}
\right]
e^{-\frac{\tau q^{2}_r}{2m}},
\eea
where ${\bm p}_r^{}$ and ${\bm q}_r^{}$ are incoming and outgoing relative momenta.
By diagonalizing this expression, we may identify the eigenvalues 
of ${\mathcal T}_2^{}$ with $e^{-\tau E_{}^{}}$, where $E_{}^{}$ are the 
two-particle eigenvalues of the Hamiltonian we are implicitly defining. 

\begin{table}[b]
\begin{center}
\caption{\label{Table:Cn}
Results of fitting the coefficients $C^{}_{n}$
(see Sec.~\ref{ImprovedT})
to the low-energy spectrum of the two-body problem at resonance, in a box of side $N^{}_x = 16$,
for an imaginary time step $\tau = 0.05$, in lattice units.
}
\begin{tabularx}{\columnwidth}{@{\extracolsep{\fill}}c c c c c c}
       \hline
       $N^{}_\mathcal{O}$ & $C^{}_0$ & $C^{}_1$ & $C^{}_2$ & $C^{}_3$ & $C^{}_4$\\
       \hline
       \hline
1   &  0.68419    &  --   &  --    &  --   &  --  \\
2   &  0.53153    & 0.07896     & --    & --    & --     \\
3   &  0.49278    & 0.04366    & 0.01807    & --    & --     \\
4   &  0.47217     & 0.03711   & 0.00784    & 0.00467     & --     \\
5   &  0.45853     & 0.03331   &  0.00718     & 0.00132    & 0.00129    \\
\hline
\end{tabularx}
\end{center}
\end{table}

One may then tune the $C_{n}^{}$ such that 
this energy spectrum matches the one required by L\"uscher's formula for the lowest $N^{}_{\mathcal O}$ 
eigenvalues, given the desired values of the scattering parameters in Eq.~(\ref{EffectiveRangeExpansion}),
and using $E = p^2/m = \eta^2 (2\pi)^2/(m L^2)$.
Specifically, if we are interested in describing the unitary limit, the eigenvalues are 
determined by the zeros of the function $S(\eta)$ in Eq.~(\ref{Seta}). The actual fitting of the $C_{n}^{}$ 
can be performed iteratively, as described in detail in Ref.~\cite{Endresetal}. Notice, in particular,
that our expression for $\mathcal T_2^{}$ ceases to be Hermitian when we promote $A$ to be an operator. 
Therefore, one needs to use $\mathcal T_2^{\dagger}\mathcal T_2^{}$ rather than $\mathcal T_2^{}$ to 
diagonalize and fit the eigenvalues, as well as for the actual Monte Carlo calculations.

\begin{figure}[h]
\includegraphics[width=1.03\columnwidth]{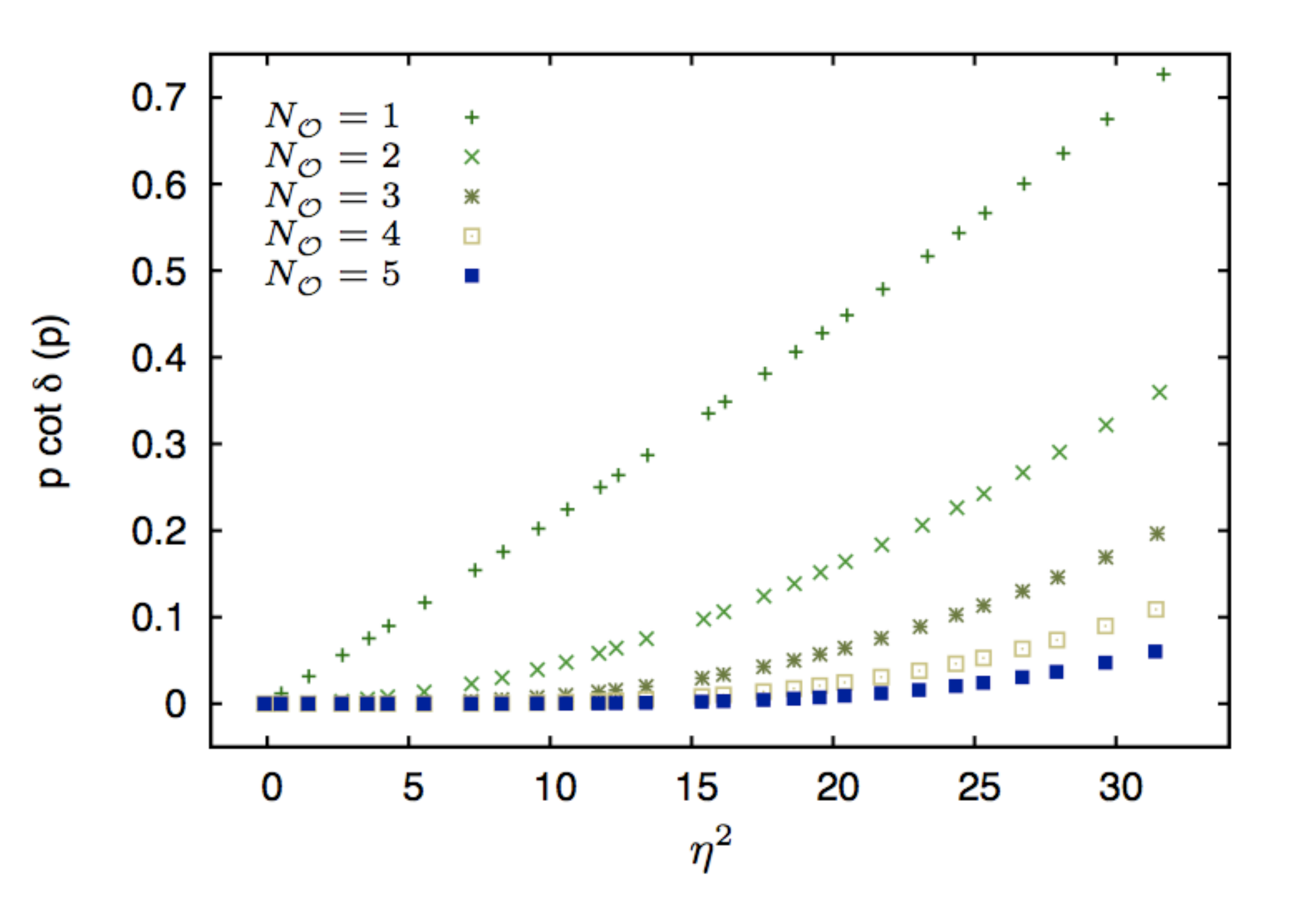}
\caption{\label{Fig:pcotd}(Color online) Plot of $p\cot \delta (p)$, in lattice units, as a function of 
$\eta^2 = {E^{}_2/p_0^2}$ (where $p_0^{} \equiv 2\pi/L$), for $N_x = 20$, $\tau = 0.05$, 
and levels of improvement $N^{}_\mathcal{O} = 1 - 5$, for a Galilean non-invariant definition of the transfer matrix 
(for $N^{}_\mathcal{O} > 2$).}
\end{figure}

The results of our fits for $C_{n}$ are shown in Table~\ref{Table:Cn}.
The quality of the improvement can be assessed by plotting $p \cot \delta(p)$ as a function of $\eta^2$,
which is shown in Fig.~\ref{Fig:pcotd} for $N_x = 20$, $\tau = 0.05$, and $N^{}_\mathcal{O} = 1 - 5$.
As can be appreciated in the figure, the same effect is achieved as in Ref.~\cite{Endresetal}: as the order
of the expansion is increased, the transfer matrix is accurately tuned to unitarity up to 
progressively higher momenta.
%

\section{\label{ImprovedO}Improved observables}

The above procedure represents a significant step forward in mitigating lattice-spacing effects 
in MC calculations, especially considering that it requires only a small coding investment for its 
implementation in extant MC codes, and it results in minimal computational overhead.

A somewhat unsettling issue remains, however, particularly in connection with improving finite temperature 
lattice calculations, such as those of Refs.~\cite{BDM, DLT, DLWM}. Indeed, in those calculations,
as well as in similar ground-state approaches, the transfer matrix is not the only object carrying 
lattice-spacing effects: the operators used to compute
expectation values also suffer from the same problems. The situation may appear problematic at first sight,
as the improvement strategy defined above does not directly define improved operators that we 
could use to compute observables at finite temperature.

On the other hand, in conventional formulations, knowledge of the explicit $\tau$ dependence of the 
transfer matrix (as in the simple example described in Sec.~\ref{SimpleExample}) allows us to take a 
derivative that brings the Hamiltonian down from the exponent (c.f. Eqs.~(\ref{T_HS1}) and (\ref{T_HS2})), 
which results in a practical expression for the calculation of the energy~\footnote{
This is, of course, just a simplified case of the generating functional formalism (see e.g. Ref.~\cite{ZinnJustinbook}), 
where one introduces a source coupled to the operator of interest, and then takes functional derivatives to bring down 
the desired powers of the operator from the exponent. In this example, the ``source" is of course just the time 
step $\tau$, or equivalently the inverse temperature parameter $\beta$.
}.

To fix ideas, let us consider the grand canonical ensemble (see also Ref.~\cite{BDM, BSS}), where the
partition function satisfies, by definition,
\beq
\mathcal Z = e^{-\beta \Omega}.
\eeq
Averages of observables can be extracted from knowledge of $\mathcal Z$ through derivatives,
and in particular the energy is obtained by means of
\beq
\label{dZdbeta}
-\frac{\partial \log \mathcal Z}{\partial \beta} = E - \mu N.
\eeq

When using the formalism derived in Sec.~\ref{SimpleExample}, we have 
(assuming the system is unpolarized)
\beq
\mathcal Z = 
\Tr \left [\mathcal T^{N_{\tau}} \right]
= \int \mathcal D \sigma \det 
\left [
1 + 
{\mathcal U}[\sigma]
\right]^2,
\eeq
where
\bea
\mathcal T^{N_{\tau}} &=&  \int \mathcal D \sigma\ 
\prod_t
\mathcal T^{}_{\uparrow}[ \sigma^{}_t] \mathcal T^{}_{\downarrow}[ \sigma^{}_t],\\
\beta &=& \tau N_\tau^{},
\eea
and where now $\sigma$ is to be regarded as a space-time varying field, and $\sigma^{}_t$ is $\sigma$
restricted to the $t$-th imaginary-time slice. The determinant is to be taken in the subspace of one-particle states, and
\beq
{\mathcal U}[\sigma]
\equiv 
\prod_t \mathcal T^{}_{\uparrow}[ \sigma^{}_t].
\eeq
As the derivative in Eq.~(\ref{dZdbeta}) becomes now a derivative with respect to $\tau$, it is clear that
all we require is knowing how to differentiate $\mathcal T^{}_{s}$ with respect to $\tau$, because
\beq
-\frac{\partial \log \mathcal Z}{\partial \tau} = 
\frac{1}{\mathcal Z}
\int \mathcal D \sigma
\det \left [
1 + 
{\mathcal U}
\right]^2
\Tr\left [\frac{\partial \mathcal U/\partial \tau}{1 + 
{\mathcal U}} \right],
\eeq
where
\beq
\frac{\partial \mathcal U}{\partial \tau} = 
\sum_{t_0} 
\prod_{t > t_0 } \mathcal T^{}_{\uparrow}[ \sigma^{}_t]
\frac{\partial \mathcal T^{}_{\uparrow}[ \sigma^{}_{t_0}]}{\partial \tau}
\prod_{t < t_0 } \mathcal T^{}_{\uparrow}[ \sigma^{}_t].
\eeq

We shall take the point of view that differentiation of partition functions with respect to parameters in
the Hamiltonian is the proper way to obtain expectation values of operators. We shall then generalize the 
improvement procedure of Ref.~\cite{Endresetal} to design highly improved operators (in particular for the energy 
and the contact) and accomplish this by taking formal derivatives of the transfer matrix and writing them 
in an operator expansion, fitting the expansion coefficients to reproduce the low end of the exact two-particle 
spectra. By "highly improved" we mean improvements that go well beyond tuning the first two coefficients in the
effective-range expansion.

While in the above example we have focused on the calculation of the energy, which we resume in the next section, similar derivations apply for the calculation of the contact, as we shall see in Sec.~\ref{SubS:Contact}.

\subsection{\label{SubS:Energy}Energy}

In the simple case presented in Sec.~\ref{SimpleExample}, the calculation of the first $\tau$ derivative 
results in the following expression:
\beq
-\frac{\partial \mathcal T^{}_{s}[\sigma]}{\partial \tau} = 
e^{-\frac{\tau \hat T^{}_{s}}{2}}
(\hat K + \hat U_\tau^{})
e^{-\frac{\tau \hat T^{}_{s}}{2}}
\eeq
where $\hat K$ is the following anticommutator
\beq
\label{KandU1}
\hat K \equiv \left\{\frac{\hat T^{}_{s}}{2}, (1 +  \sqrt{A}\; \sum_{\bm{i}} \hat n^{}_{s,\bm{i}} \sin \sigma_{\bm{i}})\right\},
\eeq
and
\beq
\label{KandU2}
\hat U_\tau^{} \equiv -\frac{\partial \sqrt{A}}{\partial \tau}\; \sum_{\bm{i}} \hat n^{}_{s,\bm{i}} \sin \sigma_{\bm{i}}.
\eeq
%
%

The derivative of $\sqrt{A}$ with respect to $\tau$ appearing in Eq.~(\ref{KandU2}) is known analytically in this case.
When using improved transfer matrices, however, that derivative is somewhat harder to compute, 
as we only know the coefficients $C^{}_n$ in Eq.~(\ref{Aexpansion}) as a result of a numerical fit,
which complicates the calculation of $\partial C_{n}/\partial \tau$.

Here we extend the procedure of Endres et al. to fitting the coefficients 
in the expansion of the derivative of $A({\bf p})$. There are at least two advantages in following this route over
computing the derivatives via finite differences. In the first place, the fitting procedure 
can provide a much more accurate determination of the expansion coefficients of the derivative, 
with less effort. (That being said, finite differences do provide a useful starting point for the
fitting routine.) Secondly, this route allows one to use different operators for different observables,
regardless of what is used for $A({\bf p})$ itself. This feature can potentially be very convenient: different observables 
are in general sensitive to different regions of momentum space, such that there is no a priori reason why
a given set of operators $\mathcal O_n$ should be universally useful.

The main idea behind the method remains the same as in Ref.~\cite{Endresetal}. L\"uscher's formula provides 
us with the exact two-particle spectrum, such that the eigenvalues of the exact two-particle transfer matrix and 
its $\tau$ derivative(s) are known:
\beq
\label{dTdtauexact}
-\frac{\partial \mathcal T^\text{exact}_{2}}{\partial \tau} = E_2^{} \exp\left(- \tau E_2^{} \right)
\eeq
where $E_2^{}$ are the exact two-particle energies in a continuous box~\footnote{
We use $\mathcal T^{}_{2}$ here for didactical purposes, but one may use the symmetric expression 
$\mathcal T^{\dagger}_{2}\mathcal T^{}_{2}$ instead, with straightforward modifications, as long as this 
same expression is used in the actual Monte Carlo calculations for a double step in the imaginary-time direction.
}.

In order to match this spectrum, we take the derivative of Eq.~(\ref{2particleTmatrixCOM}) with respect to $\tau$, 
evaluated between eigenstates $| E_{}^{} \rangle$ of the proposed transfer matrix $\mathcal T^{}_{2}$. 
We thus obtain, using the Feynman-Hellmann theorem,
\beq
\label{dTdtauCOM}
-\frac{\partial \langle E_{}^{} | \mathcal T^{}_{2}| E_{}^{} \rangle}{\partial \tau} = 
\langle E_{}^{} | 
e^{-\frac{\tau p^{2}_r}{2m}}
\left [
K_2^{} + U_2^{}
\right]
e^{-\frac{\tau q^{2}_r}{2m}} 
| E_{}^{} \rangle,
\eeq
where
\beq
K_2^{} \equiv
\left [
\frac{p^{2}_r}{2m}+
\frac{q^{2}_r}{2m}
\right]
\left [
\delta_{{\bm p}_r^{} {\bm q}_r^{}}\! + \frac{A({\bm p}_r^{})}{2V}
\right]
\eeq
and
\beq
U_2^{} \equiv -\frac{1}{2V} \frac{\partial A({\bm p}_r^{})}{\partial \tau} = 
\frac{1}{2V} \sum_{n=0}^{N_\text{max}-1} D^{}_{n} {\mathcal O}_{n}^{}({\bm p}_r^{}).
\eeq

The rest of the recipe consists in taking the right-hand side of Eq.~(\ref{dTdtauCOM}) and 
fitting the coefficients $D_{n}$ such that the first 
$N^{}_\text{max}$ eigenvalues of the exact expression Eq.~(\ref{dTdtauexact}) 
(which correspond to the lowest eigenvalues prescribed by L\"uscher's formula) are reproduced.
We may assume at this point that the coefficients $C^{}_{n}$ are known, such that the eigenvectors 
$| E_{}^{} \rangle$ are fixed when we set out to find the $D^{}_{n}$. In that case, the $D^{}_{n}$ are
determined by a linear system of equations of order $N_\text{max} \times N_\text{max}$:
\beq
\sum_{n=0}^{N_\text{max}-1} M^{}_{E n} D^{}_{n} = Y^{}_{E}
\eeq
where
\beq
M^{}_{E n} = \frac{1}{2V} \langle E_{}^{} | {\mathcal O}_{n}^{}| E_{}^{} \rangle
\eeq
and
\beq
Y^{}_{E} = E \exp\left(- \tau E \right) -
\langle E_{}^{} | e^{-\frac{\tau p^{2}_r}{2m}} K_2^{} e^{-\frac{\tau q^{2}_r}{2m}} | E_{}^{} \rangle
\eeq

Once the coefficients $D_{n}$ have been determined, one can use Eq.~(\ref{KandU2}) in a lattice calculation
simply by replacing $A \to A({\bf p})$, and taking $\partial C^{}_{n}/\partial \tau = D_{n}$.
The results of the fits for $C_{n}$, $D_{n}$ are shown in Tables~\ref{Table:Cn} and~\ref{Table:Dn}.


\begin{table}[t]
\begin{center}
\caption{\label{Table:Dn}
Results of fitting the coefficients $D^{}_{n}$
(see Sec.~\ref{ImprovedO})
to the low-energy spectrum of the two-body problem at resonance, in a box of side $N^{}_x = 16$,
for an imaginary time step $\tau = 0.05$, in lattice units.
}
\begin{tabularx}{\columnwidth}{@{\extracolsep{\fill}}c c c c c c}
       \hline
       $N^{}_\mathcal{O}$ & $D^{}_0$ & $D^{}_1$ & $D^{}_2$ & $D^{}_3$ & $D^{}_4$\\
       \hline
       \hline
1   &  -14.76869    &  --   &  --   &  --   &  --  \\
2   &  -11.54894     & -1.74519     & --    & --    & --     \\
3   &  -10.74506     &  -0.96946     & -0.40164    & --    & --     \\
4   & -10.31974     & -0.82605    & -0.17494     & -0.10404   & --     \\
5   & -10.03874     & -0.74266    & -0.16064    & 0.02948    & -0.02878      \\
\hline
\end{tabularx}
\end{center}
\end{table}

\begin{figure}[t]
\includegraphics[width=1.03\columnwidth]{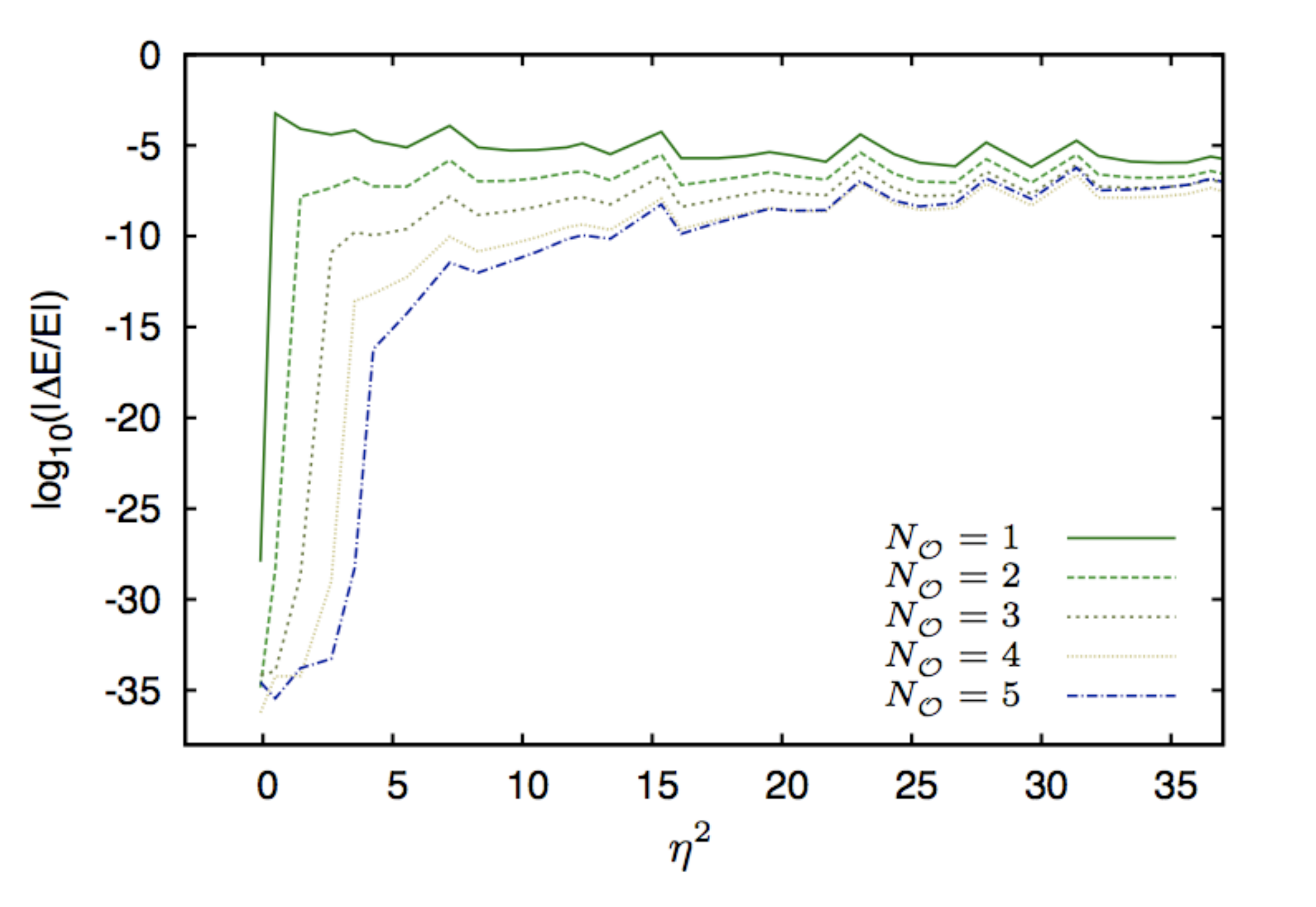}
\caption{
\label{Fig:E_Spectra}
(Color online) Logarithmic plot of the difference between the exact spectrum of the
two-body problem Eq.~(\ref{dTdtauexact}) at unitarity, and the approximate spectrum obtained with 
various levels of improvement $N^{}_\mathcal{O} = 1 - 5$, relative to the exact spectrum (see text for details),
as a function of $\eta^2 = {E^{}_2/p_0^2}$, where $p_0^{} \equiv 2\pi/L$. 
The original data are discrete; the lines are intended as a guide to the eyes.
These results correspond to a lattice of side $N_x = 20$ and temporal spacing $\tau = 0.05$.}
\end{figure}

As a first illustration of the level of improvement that can be achieved for the energy, 
Fig.~\ref{Fig:E_Spectra} shows the difference between the approximate spectrum with various levels 
of improvement $E_\text{approx}$ and the exact spectrum $E_\text{exact}$, as a function of $\eta^2$,
through the quantity $\text{Log}(|\Delta E|)$, where $\Delta E= (E_\text{approx} - E_\text{exact})/E_\text{exact}$.
As expected, with each new parameter a new eigenvalue is reproduced, with the concomitant reduction in 
the error.

Surprisingly, the approximations perform well considerably beyond the lowest eigenvalues that are fit. 
Indeed, with each new level of improvement we see, apart from a dramatic reduction in the error for the 
target eigenvalues, an extra reduction in the error that is evident even beyond $\eta^2 \simeq 30$.

\subsection{\label{SubS:Contact}Contact}

One of the many ways to define the contact $C$ (see Refs.~\cite{ShinaContact, BraatenReview}) is 
through the derivative of the energy with respect to the inverse scattering length:
\beq
\frac{\partial E}{ \partial a^{-1}_{}} = -\frac{\hbar^2}{4 \pi m} C.
\eeq
Since $E$ may be obtained directly from the logarithm of the partition function, we are again in a situation
where we require a derivative of the transfer matrix with respect to a parameter, in this case $a^{-1}_{}$.
The generalization of the above definition to finite temperature in the grand canonical ensemble is
\beq
\left(\frac{\partial \Omega}{ \partial a^{-1}_{}} \right)_{T,\mu} = 
-\frac{1}{\beta} \left(\frac{\partial \log \mathcal Z}{ \partial a^{-1}_{}} \right)_{T,\mu} 
= -\frac{\hbar^2}{4 \pi m} C,
\eeq
where $\Omega$ is the grand thermodynamic potential, $T$ is the temperature and $\mu$ is the
chemical potential.

In many-body lattice calculations, using these definitions involves the following expression:
\beq
\frac{\partial \mathcal T^{}_{s}[\sigma]}{\partial a^{-1}_{}} = 
e^{-\frac{\tau \hat T^{}_{s}}{2}}
\hat U_{a^{-1}_{}}^{}
e^{-\frac{\tau \hat T^{}_{s}}{2}}
\eeq
where
\beq
\label{UU}
\hat U_{a^{-1}_{}}^{} \equiv \frac{\partial \sqrt{A}}{\partial a^{-1}_{}}\; \sum_{\bm{i}} \hat n^{}_{s,\bm{i}} \sin \sigma_{\bm{i}}.
\eeq

The first step towards using these expressions in combination with the improvement procedure
is to take a formal derivative of $\mathcal T$ with respect to $a^{-1}_{}$ in the two-particle space, which we can 
treat exactly:
%
%
%
\beq
\label{dT2dainvexact}
\frac{\partial \mathcal T_2^\text{exact}}{\partial a^{-1}_{}} =
-\tau \frac{\partial E^{}_2}{\partial a^{-1}_{}}\exp\left(- \tau E_2^{} \right).
\eeq
In order to compute the change in the exact two-particle energy $E_2^{}$ due to a small change in the 
inverse scattering length, we use the fact that the energies are implicitly defined as solutions of 
Eq.~(\ref{Luscherformula}), which implies
\beq
\label{dE2dainv}
\frac{\partial E^{}_{2}}{\partial a^{-1}_{}} = - \frac{4\pi^3}{L}\left(\frac{d \mathcal S}{d \eta^2_{}} \right)^{-1}
\eeq
where $\eta^2 = E_2^{} L^2/(2\pi)^2$, and the derivative on the right-hand side is to be evaluated 
at the corresponding solution of Eq.~(\ref{Luscherformula}). Table~\ref{Table:etastar} in the Appendix 
shows the first 30 roots of $S(\eta)$ and the corresponding values of ${d \mathcal S}/{d \eta^2_{}}$.
In the derivation of Eq.~(\ref{dE2dainv}), we have assumed that all the effective-range parameters 
other than the scattering length are kept constant.

Having the exact target spectrum, we proceed by finding the corresponding expression in terms of 
the HS function $A({\bf p})$, which we obtain using Eq.~(\ref{2particleTmatrixCOM}) and the 
Feynman-Hellmann theorem:
\beq
\label{dTdainvCOM}
\frac{\partial \langle E | \mathcal T^{}_{2} | E \rangle }{\partial a^{-1}_{}} = 
\langle E | e^{-\frac{\tau p^{2}_r}{2m}}
\frac{1}{2V} \frac{\partial A({\bm p}_r^{})}{\partial a^{-1}_{}}
e^{-\frac{\tau q^{2}_r}{2m}} | E \rangle,
\eeq
where, as before, we expand in terms of our chosen set of operators,
\beq
\frac{\partial A({\bm p}_r^{})}{\partial a^{-1}_{}}
=
\sum_{n=0}^{N_\text{max}} F^{}_{n} {\mathcal O}_{n}^{}({\bm p}_r^{}),
\eeq
and we determine the coefficients $F^{}_{n}$ by fitting the diagonal matrix elements in the right-hand side 
of Eq.~(\ref{dTdainvCOM}) to the exact spectrum of Eqs.~(\ref{dT2dainvexact}) and (\ref{dE2dainv}).
As with the energy, the fitting procedure can be reduced to solving a set of linear equations of order 
$N_\text{max} \times N_\text{max}$.
Illustrative results of such a fit are shown in Table~\ref{Table:Fn}.


\begin{table}[t]
\begin{center}
\caption{\label{Table:Fn}
Results of fitting the coefficients $F^{}_{n}$
(see Sec.~\ref{ImprovedO})
to the low-energy spectrum of the two-body problem at resonance, in a box of side $N^{}_x = 16$,
for an imaginary time step $\tau = 0.05$, in lattice units.
}
\begin{tabularx}{\columnwidth}{@{\extracolsep{\fill}}c c c c c c}
       \hline
       $N^{}_\mathcal{O}$ & $F^{}_{0}$ & $F^{}_1$ & $F^{}_2$ & $F^{}_3$ & $F^{}_4$\\
       \hline
       \hline
1   &   0.36773   &  --   &  --   &  --   &  --  \\
2   &  0.14532   & 0.07568    & --    & --    & --     \\
3   &  0.11370   & 0.02220    & 0.01957   & --    & --     \\
4   &   0.09659   & 0.01695    & 0.00415   & 0.00538   & --     \\
5   &  0.08205   & 0.01278    & 0.00406   & -0.00023   & 0.00180     \\
\hline
\end{tabularx}
\end{center}
\end{table}
%
\begin{figure}[t]
\includegraphics[width=1.03\columnwidth]{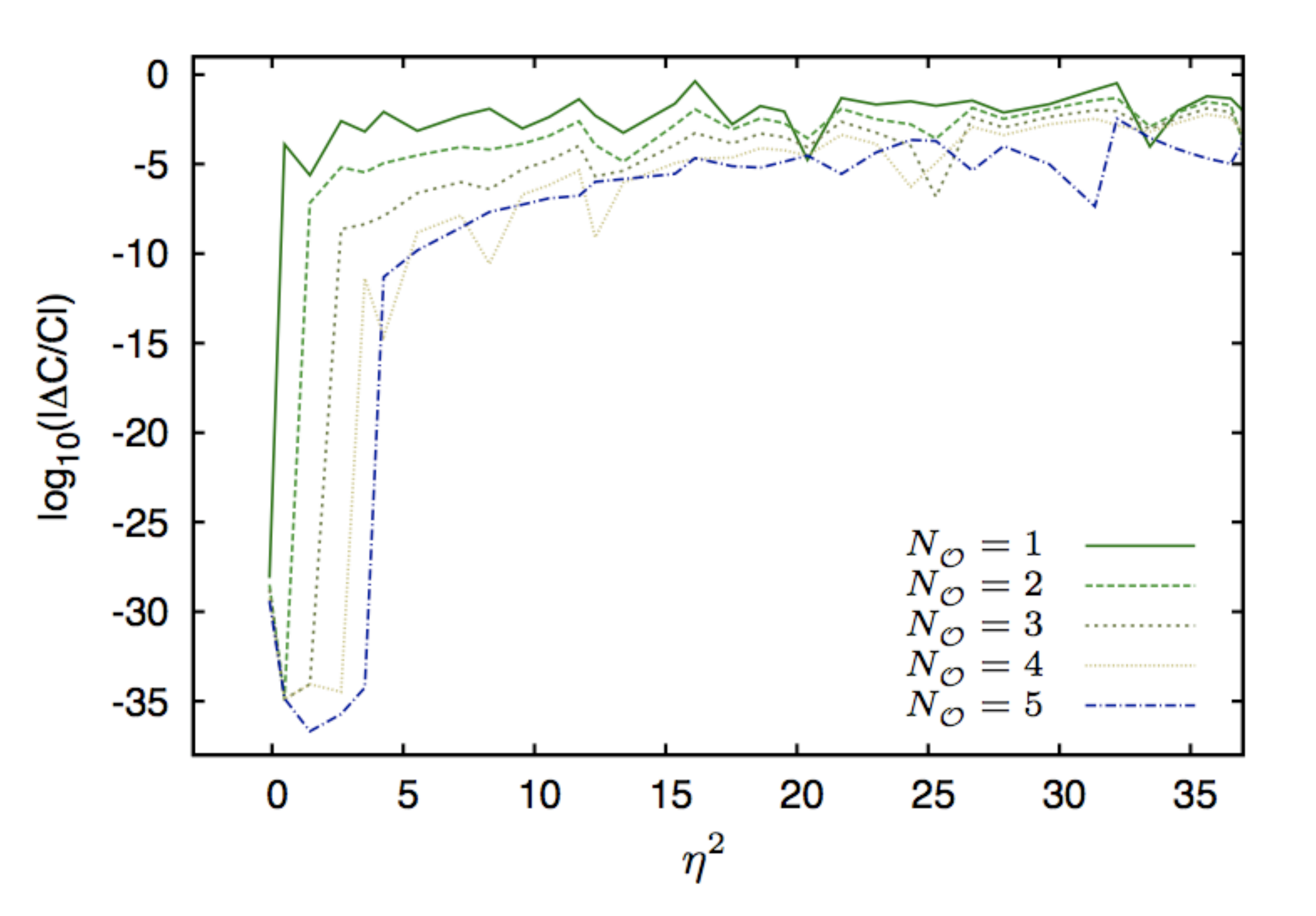}
\caption{
\label{Fig:C_Spectra}
(Color online) Logarithmic plot of the difference between the exact two-body spectrum
Eq.~(\ref{dT2dainvexact}) at unitarity, and the spectrum obtained with various levels 
of improvement $N^{}_\mathcal{O} = 1 - 5$, relative to the exact spectrum (see text for details),
as a function of $\eta^2 = {E^{}_2/p_0^2}$, where $p_0^{} \equiv 2\pi/L$. 
The original data are discrete; the lines are intended as a guide to the eyes.
These results correspond to a lattice of side $N_x = 16$ and temporal 
spacing $\tau = 0.05$.}
\end{figure}
As with the energy, a first glimpse at the level of improvement that can be obtained at this point. 
This is shown in Fig.~\ref{Fig:C_Spectra}, where we display the difference $\Delta C$ 
between the spectrum with various levels of improvement and the exact spectrum, 
divided by the latter.
As in Fig.~\ref{Fig:E_Spectra}, each new parameter allows one to fit a new eigenvalue to high accuracy, 
matching the desired physics beyond the lowest momentum.

Unlike in Fig.~\ref{Fig:E_Spectra}, the improvement for eigenvalues beyond those 
explicitly fit is limited, breaking down after the eighth or ninth eigenvalue. From that point on toward 
higher energies, little improvement, if any, is observed as the order of the expansion is increased.
This behavior is only unexpected in the light of Fig.~\ref{Fig:E_Spectra}, where the situation is 
(surprisingly) much more favorable. Possibly, part of the reason for this behavior is that the target spectra
for the energy and the transfer matrix are monotonic (either increasing or decreasing) at low energies, 
whereas the one for the contact is closer to a random sequence (see Table~\ref{Table:etastar} in
the Appendix).

On the other hand, as mentioned before, there is no reason to believe that a given choice of operators 
will be equally useful for all observables. It is therefore natural to suspect that a different set of operators 
could yield a better expansion for the contact. This possibility remains to be studied. 

\subsection{{\label{SubS:Extrapolation}}Extrapolation to the ground state}

Taking a Slater-determinant state $| \psi_0 \rangle$ as a starting point, it is easy to see that the probability sum in 
ground-state calculations, for a temporal extent $\beta$ can be written as
\beq
\mathcal Z^{}_0(\beta) = \sum_k A^{}_k e^{-\beta E_k^{}},
\eeq
where $E_k^{}$ are the exact energy eigenvalues and
\beq
A^{}_k \equiv |\langle \psi_0 | E_k^{} \rangle |^2.
\eeq
Taking a derivative of $\log \mathcal Z^{}_0(\beta)$ with respect to $\tau$ it is easy to see that
in the large-$\beta$ limit one can write
\beq
E(\beta) \equiv -\frac{\partial \log \mathcal Z^{}_0(\beta)}{\partial \beta} 
\to E_0^{} + b_E^{} e^{-\beta \delta},
\eeq
where $E_0$ is the ground-state energy, $\delta = E_1 - E_0$, and
\beq
b_E^{} = \frac{A_1^{}}{A_0^{}}(E^{}_1 - E^{}_0).
\eeq

Similarly, taking a derivative with respect to $a^{-1}$ one can see that the first few leading contributions to 
the asymptotic behavior at large $\beta$ are given by
\beq
C(\beta) \equiv \frac{4 \pi m}{\hbar^2 \beta}\frac{\partial \log \mathcal Z^{}_0(\beta)}{\partial a^{-1}_{}} 
\to C_0^{} + b_{C1}^{}\beta^{-1} + b_{C2}^{} e^{-\beta \delta}
\eeq
where $C_0$ is the ground-state contact,
\bea
b_{C1}^{} &=& \frac{4 \pi m}{\hbar^2} \frac{\partial \log A_0^{}}{\partial a^{-1}_{}},
\\
b_{C2}^{} &=& -\frac{4 \pi m}{\hbar^2}\frac{A_1^{}}{A_0^{}}
\left( \frac{\partial E^{}_1}{\partial a^{-1}_{}} - \frac{\partial E^{}_0}{\partial a^{-1}_{}} \right).
\eea
We shall use these expressions to motivate the extrapolations to the $\beta \to \infty$ 
limit below.

\section{\label{Results}Illustrative results and conclusions}

\begin{figure}[h]
\includegraphics[width=1.03\columnwidth]{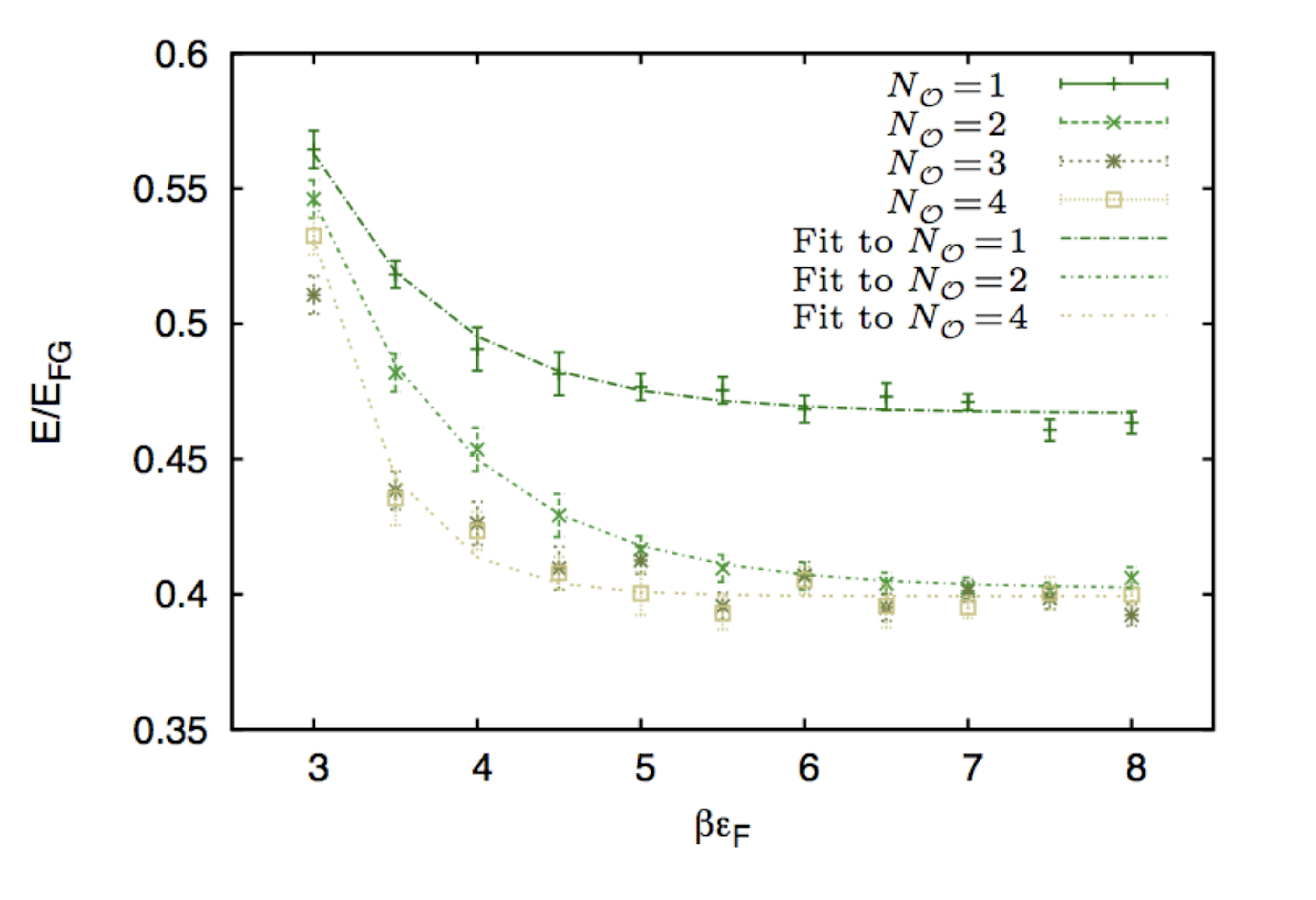}
\caption{
\label{Fig:EGS_40}
(Color online) Energy, in units of $E^{}_{FG} = \frac{3}{5} N \epsilon_F^{}$, as a function of the 
extent of the time direction $\beta \epsilon_F^{} = \tau N_\tau \epsilon_F^{}$,
for 80 particles and levels of improvement $N^{}_\mathcal{O} = 1 - 4$.}
\end{figure}

\begin{figure}[h]
\includegraphics[width=1.03\columnwidth]{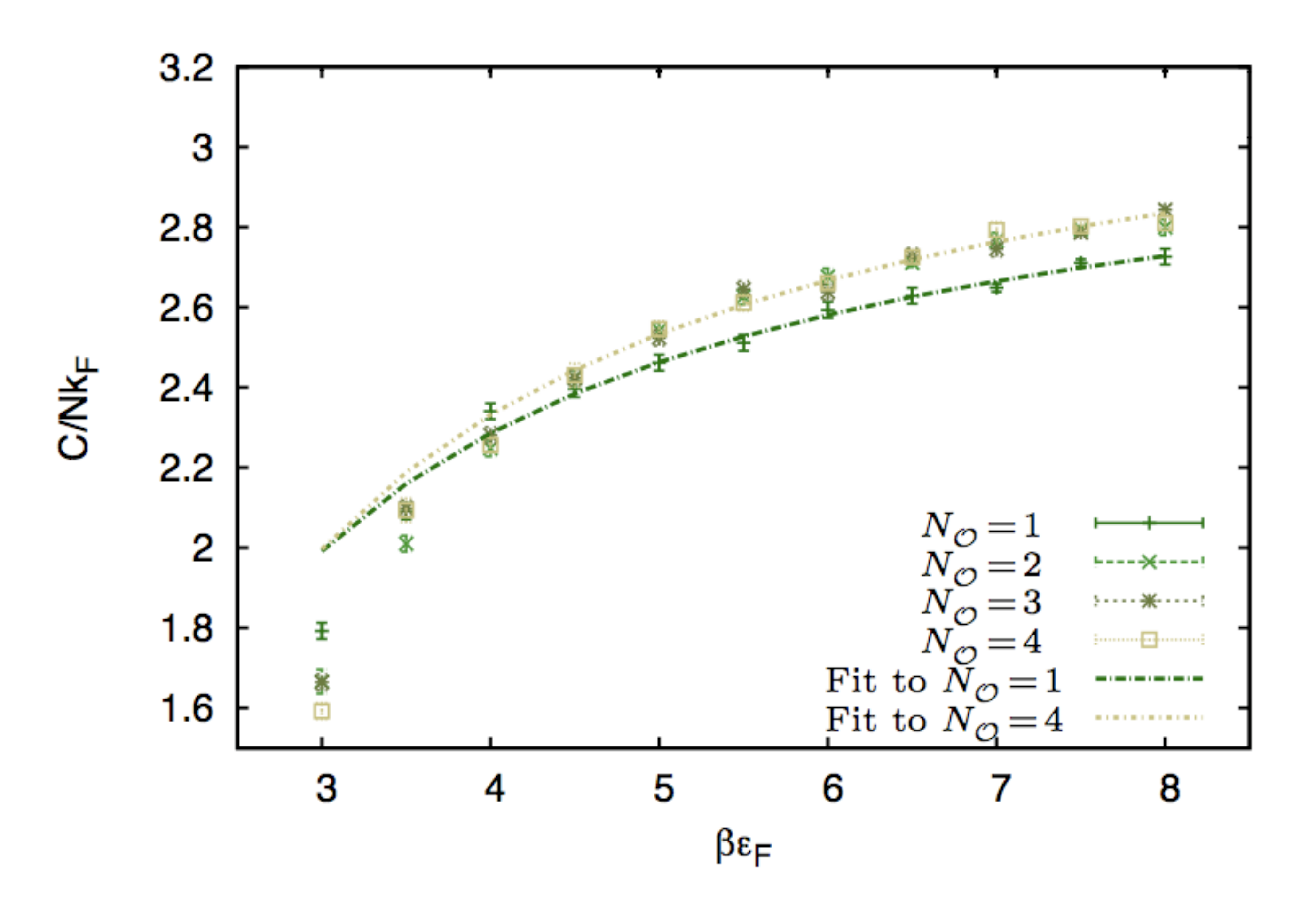}
\caption{
\label{Fig:Contact_40}
(Color online) Contact, in units of $N k_F^{}$, where $k_F^{} = (3 \pi^2 N/V)^{1/3}$ is the Fermi momentum, 
as a function of the extent of the time direction $\beta \epsilon_F^{} = \tau N_\tau \epsilon_F^{}$,
for 80 particles and levels of improvement $N^{}_\mathcal{O} = 1 - 4$. Also shown are fits to the $N^{}_\mathcal{O}\!=\!1$ and
$N^{}_\mathcal{O}\!=\!4$ datasets using the first two terms in the asymptotic form of the previous section (see text for details).}
\end{figure}

To illustrate the effect of improved operators in realistic lattice calculations, this section
presents the results of ground-state MC calculations of the energy (Fig.~\ref{Fig:EGS_40}) and 
the contact (Fig.~\ref{Fig:Contact_40}), for an unpolarized system at unitarity.

Results are shown for 80 particles (40 per spin) in a volume of $10^3$ lattice points, for 
various levels of improvement $N^{}_\mathcal{O} = 1 - 4$.
In each case, calculations were performed for time directions of extent 
$\beta \epsilon_F^{} = 2.0 - 8.0$ (corresponding to $N^{}_{\tau}$ roughly between 40 and 200), 
subsequently extrapolating to the $\beta \to \infty$ limit. 
We have taken $\tau = 0.05$ in lattice units, and a Slater determinant of plane 
waves as the starting guess for the ground-state wavefunction. For each value of $\beta$, we obtained 
approximately $400$ samples of the auxiliary field $\sigma$. The statistics is enhanced by 
a factor of $20-100$ by the fact that the operators we have defined utilize every other time slice.
Some obvious fluctuations remain, as evident from some degree of jaggedness in the data.
This can be resolved by increasing the statistics, but it does not affect our main conclusions.


Once the improvements are turned on, i.e. for $N_{\mathcal O} = 2-4$, we see that the change from
$k_F^{} r^{}_\text{eff} \simeq 0.54$ to $k_F^{} r^{}_\text{eff} = 0$ results in a considerable reduction in the energy. 
Our extrapolations to the ground state yield $\xi \equiv E/E^{}_{FG} = 0.467(2)$ for the unimproved case, and 
$\xi = 0.402(1), 0.399(2), 0.399(2)$ for $N_{\mathcal O} = 2,3,4$, respectively. This change, of roughly $15\%$, is in line 
with our expectations based on the large-scale calculations of Ref.~\cite{Carlsonetal}. While these results are still roughly $5\%$ 
greater than those of Ref.~\cite{Carlsonetal}, which found $\xi \approx 0.38$, it is remarkable that this can be achieved with a 
$10^3$ lattice. The remaining effects must be a combination of finite volume and finite imaginary-time step $\tau$ effects.

A similar situation is observed for the contact. Using the extrapolation formula of the previous section at leading plus next-to leading order, 
we obtain $C/(N k_F^{}) = 3.17(3)$ in the unimproved case, and $C/(N k_F^{}) = 3.31(4), 3.34(4)$ and $3.30(4)$
for $N_{\mathcal O} = 2,3,4$, respectively. (We have discarded three data points at the lowest values of $\beta \epsilon_F^{}$ 
for the purpose of capturing the asymptotic behavior at large $\beta$; the fits are stable against further removal of points at low 
$\beta \epsilon_F^{}$). 
While the latter extrapolations clearly overlap when taking the uncertainty into account, they do not overlap with the unimproved case,
which is clearly consistent with what we see in Fig.~\ref{Fig:Contact_40} at large $\beta$. 
The remaining systematic errors, likely due to volume effects for the most part, appear to be only as large as 
$3\%$, if we consider the most recent estimates of the contact in the ground state ($\approx 3.39$, see Ref.~\cite{Stefano}). As in the case of
the energy, it is remarkable that such a small volume as $10^3$ already yields a result that is quite close to the best current estimate.

In conclusion, we have presented a methodology based on the work of Refs.~\cite{Endresetal, Endresetal2}, whereby one can
generate not only improved transfer matrices but also improved operators, which account for finite-range effects in a systematic
fashion. The improvement program can be carried out in a Galilean invariant or in a Galilean non-invariant way. We have chosen the latter
because it provides a better approach to the unitary limit. As shown in Appendix~\ref{AppendixGalileanInvariance}, 
the difference in the eigenvalues of the invariant and non-invariant transfer matrices is very small, such that the effect
can be safely considered to be negligible. This is likely due to the fact that the lattice already breaks Galilean invariance.
With the chosen improvements in place, we see a large change for the energy but a rather small change for the contact.
Indeed, once finite-range effects are under control, the dominant contribution to systematic uncertainties is expected to be 
given by finite-volume effects. If so, the latter appear to be relatively small both for the energy (roughly $5\%$) as well as 
for the contact (roughly $3\%$).

While we have focused on the unitary limit, the tuning procedure for the transfer matrix and the operators can easily be applied to systems
away from unitarity as well as away from the zero effective-range limit. In combination with hybrid Monte Carlo (HMC) techniques, which
have recently made it possible to access larger volumes than ever before (as large as $20^3$), we expect the method presented here
to provide a powerful strategy to tackle the non-relativistic many-fermion problem. This applies in particular to finite-temperature
calculations where lowering the ultraviolet cutoff (while keeping effective-range effects under control) reduces the size of the basis, speeding up 
the computations considerably. 

In this regard, it should also be pointed out that improving the transfer matrix affects the performance of the HMC algorithm only marginally.
Indeed, the scaling with system size (particle number, spacetime volume) remains unchanged when increasing $N_{\mathcal O}$. Only 
the prefactor in the scaling law increases somewhat (by about $20-25\%$) due to the need for extra Fourier transforms when applying 
the operator $A(p)$. Aside from this, the force in the molecular dynamics (MD) part of the HMC algorithm becomes somewhat larger when 
$A(p)$ is improved, such that somewhat smaller MD time steps (not to be confused with $\tau$) are needed. In short, the performance
of the HMC algorithm is largely unaffected by the use of improved transfer matrices.

Finally, we would like to stress that the operator proposed here for the calculation of the contact (based on the derivative of the 
two-body energy, in turn computed using Eq.~(\ref{dE2dainv})) constitutes a novel way to calculate $C$ in MC calculations. This 
method should be contrasted with others based on extracting the derivative of the equation of state via finite differences, using the 
momentum distribution, or using the pair distribution function. Each of these will, in general, behave differently in terms of their 
systematic effects.

\acknowledgements

Thanks are in order to J. Carlson, W. Detmold, T. L\"ahde, D. Lee, Y. Nishida, E. Passemar, and S. Tan 
for discussions and/or critical comments on the manuscript. I am especially indebted to E. R. Anderson, for numerous 
discussions that led to this work, and to A. N. Nicholson for clarifications on the work of Ref.~\cite{Endresetal}. 
I would also like to thank the Institute for Nuclear Theory in Seattle, WA, as well as 
the ``Quarks, Hadrons and Nuclei" group at the University of Maryland, where some of the early work and 
discussions took place.
This work was supported in part by U.S. Department of Energy, Office of Nuclear Physics under contract
DE-FC02-07ER41457 (UNEDF SciDAC).

\appendix
\section{\label{AppendixSeta} Evaluation of $\mathcal S(\eta)$.}

There are many ways to evaluate the function $\mathcal S(\eta)$ efficiently. The one that I have found easiest to 
implement and understand is essentially the one communicated to me by Shina Tan, which I reproduce here,
with some minor modifications that I have found useful.

The first step is to enhance the convergence of the sum by introducing an exponential factor and separating
the singularity in the tail (see Ref.~\cite{Shina}):
\beq
\sum_{|{\bf n}| < \Lambda} \frac{1}{n^2 - \eta^2} = 
\sum_{|{\bf n}| < \Lambda} \frac{1 - e^{-x(n^2 - \eta^2)}}{n^2 - \eta^2} + 
\sum_{|{\bf n}| < \Lambda} \frac{e^{-x(n^2 - \eta^2)}}{n^2 - \eta^2},
\eeq
where $n = |\bf{n}|$, the sums are over all triplets of integers $\bf{n}$, and $x$ is a small positive parameter. 
The second sum has no convergence problems, in fact it converges very quickly, so we shall put 
it aside by defining 
\beq
\tilde{\mathcal S}(\eta,x) \equiv
\lim_{\Lambda \to \infty} \sum_{|{\bf n}| < \Lambda} \frac{e^{-x(n^2 - \eta^2)}}{n^2 - \eta^2}.
\eeq
%

\begin{table}[t]
\begin{center}
\caption{\label{Table:etastar}
First 30 roots of $\mathcal S(\eta)$, and ${d \mathcal S}/{d \eta^2}$ evaluated at those roots.
}
\begin{tabularx}{\columnwidth}{@{\extracolsep{\fill}}c c c}
       \hline
       $k$ & ${\eta^2_k}$ & $d \mathcal S/d{\eta^2_k}$ \\
       \hline
       \hline
1   &   -0.0959007    &   123.82387   \\ 
2   &    0.4728943    &    39.75514   \\ 
3   &    1.4415913    &    82.36519   \\ 
4   &    2.6270076    &   106.24712   \\ 
5   &    3.5366199    &    84.23133   \\ 
6   &    4.2517060    &   161.88763   \\ 
7   &    5.5377008    &   212.49220   \\ 
8   &    7.1962632    &    62.95336   \\ 
9   &    8.2879537    &   231.79580   \\ 
10   &    9.5345314    &   247.82611   \\ 
11   &   10.5505341    &   233.82976   \\ 
12   &   11.7014957    &   185.61411   \\ 
13   &   12.3102392    &   183.65019   \\ 
14   &   13.3831152    &   316.68684   \\ 
15   &   15.3537375    &    82.86757   \\ 
16   &   16.1218253    &   506.59914   \\ 
17   &   17.5325415    &   371.40245   \\ 
18   &   18.6053932    &   308.00372   \\ 
19   &   19.5186394    &   255.97969   \\ 
20   &   20.4033187    &   329.98905   \\ 
21   &   21.6944179    &   394.81924   \\ 
22   &   23.0194727    &    94.98929   \\ 
23   &   24.3306210    &   342.25749   \\ 
24   &   25.3016129    &   526.27127   \\ 
25   &   26.6803600    &   514.90705   \\ 
26   &   27.8780019    &   150.20773   \\ 
27   &   29.6156511    &   548.38017   \\ 
28   &   31.3536974    &   114.02114   \\ 
29   &   32.1958982    &   443.21169   \\ 
30   &   33.4483351    &   452.78989   \\

\hline
\end{tabularx}
\end{center}
\end{table}


The first sum, on the other hand, converges just as slowly as the original one
(once we subtract the $4\pi \Lambda$ term). Let us then analyze this first sum. For  $0 < x \ll 1$,
we may approximate it very accurately in terms of an integral:
\beq
\sum_{|{\bf n}| < \Lambda} \frac{1 - e^{-x(n^2 - \eta^2)}}{n^2 - \eta^2} 
\simeq
4\pi \int_0^{\Lambda}\!\!\!\! dn\ \frac{n^2 \left (1 - e^{-x(n^2 - \eta^2)} \right)}{n^2 - \eta^2},
\eeq
where we have used the spherical symmetry of the integrand. Writing
\beq
\frac{n^2}{n^2 - \eta^2} = 1 + \frac{\eta^2}{n^2 - \eta^2},
\eeq
we separate the integrals into two terms. First,
\beq
4\pi \int_0^{\Lambda}\!\!\!\! dn \left(1 - e^{-x(n^2 - \eta^2)} \right) = 
4\pi \Lambda - 4\pi e^{x\eta^2}\!\!\! \int_0^{\Lambda}\!\!\!\! dn \ e^{-x n^2},
\eeq
where the $4\pi \Lambda$ cancels with the $-4\pi \Lambda$ term in Eq.~(\ref{Seta}),
and the integral in the second term is equal to $1/2\sqrt{{\pi}/{x}}$ in the limit $\Lambda \to \infty$. 
Second, we have
\beq
\mathcal I \equiv 4\pi \int_0^{\Lambda}\!\!\!\! dn \frac{\eta^2 \left (1 - e^{-x(n^2 - \eta^2)} \right)}{n^2 - \eta^2}.
\eeq
To treat this term we write
\beq
1 - e^{-x(n^2 - \eta^2)} = x (n^2 - \eta^2) \int_0^1\!\!\!\! dy \ e^{-x y (n^2 - \eta^2)} 
\eeq
and exchange the order of integration of $y$ and $n$, to obtain
\beq
\mathcal I = 4\pi x \eta^2 \int_0^1\!\!\!\! dy  \ e^{x y \eta^2}  \int_0^{\Lambda}\!\!\!\! dn \ e^{-x y n^2},
\eeq
all of which can be easily evaluated in the limit $\Lambda \to \infty$, where the full result can be written as
\beq
\mathcal S(\eta) = 
\tilde{\mathcal S}(\eta,x) - 
\frac{2 \pi^{3/2}}{\sqrt{x}}
\left( 
e^{x \eta^2}
- 2 x \eta^2 \!\!\int_0^{1}\!\!\! dt\ e^{x \eta^2 t^2}
\right),
\eeq
which is to be evaluated for $0 < x \ll 1$.

With the above expressions at hand, it is straightforward to find the first derivative of $\mathcal S$ with respect to $\eta^2$:
\beq
\frac{d \mathcal S}{d \eta^2} = \frac{d \tilde{\mathcal S}}{d \eta^2} -
2 \pi^{3/2} 
\sqrt{x} \left( 
e^{x \eta^2}
 -
2 \int_0^{1}\!\!\!\! dt\ e^{x \eta^2 t^2}(1 + x \eta^2 t^2)
\right),
\eeq
where
\beq
 \frac{d \tilde{\mathcal S}}{d \eta^2} = 
 x \tilde{\mathcal S}(\eta) + 
\lim_{\Lambda \to \infty} 
\sum_{|{\bf n}| < \Lambda} \frac{e^{-x(n^2 - \eta^2)}}{(n^2 - \eta^2)^2}.
\eeq

Table~\ref{Table:etastar} shows the first 30 roots of $\mathcal S(\eta)$, along with 
the corresponding values of the derivative ${d \mathcal S}/{d \eta^2}$.


\section{\label{AppendixGalileanInvariance} Galilean invariance vs. non-invariance.}

One cannot fail to notice that not only is the transfer matrix of Eq.~(\ref{2particleTmatrix}) not symmetric, an issue
we managed to deal with above, it is also not fully Galilean invariant. Indeed, while translation and rotation invariance
are preserved (in their discrete forms set by the lattice), Galilean-boost symmetry is broken. This is because we 
have assumed that $A$ was promoted to an operator that acts on everything that appears to its right. 

On the other hand, if we assume that $A$ acts on the auxiliary field only, as in Ref.~\cite{Endresetal}, one arrives 
directly at a local, Galilean-invariant lattice theory. 
Indeed, in that case Eq.~(\ref{2particleTmatrix}) changes form in a very simple manner, namely the function 
$A({\bm p}_\uparrow^{})$ is replaced by $A({\bm p}_\uparrow^{} \!-\! {\bm q}_\uparrow^{})$, and similarly for the other spin.
The function $A$ has then a clear physical significance as the momentum transfer in a two-body collision.
\begin{figure}[h]
\includegraphics[width=1.03\columnwidth]{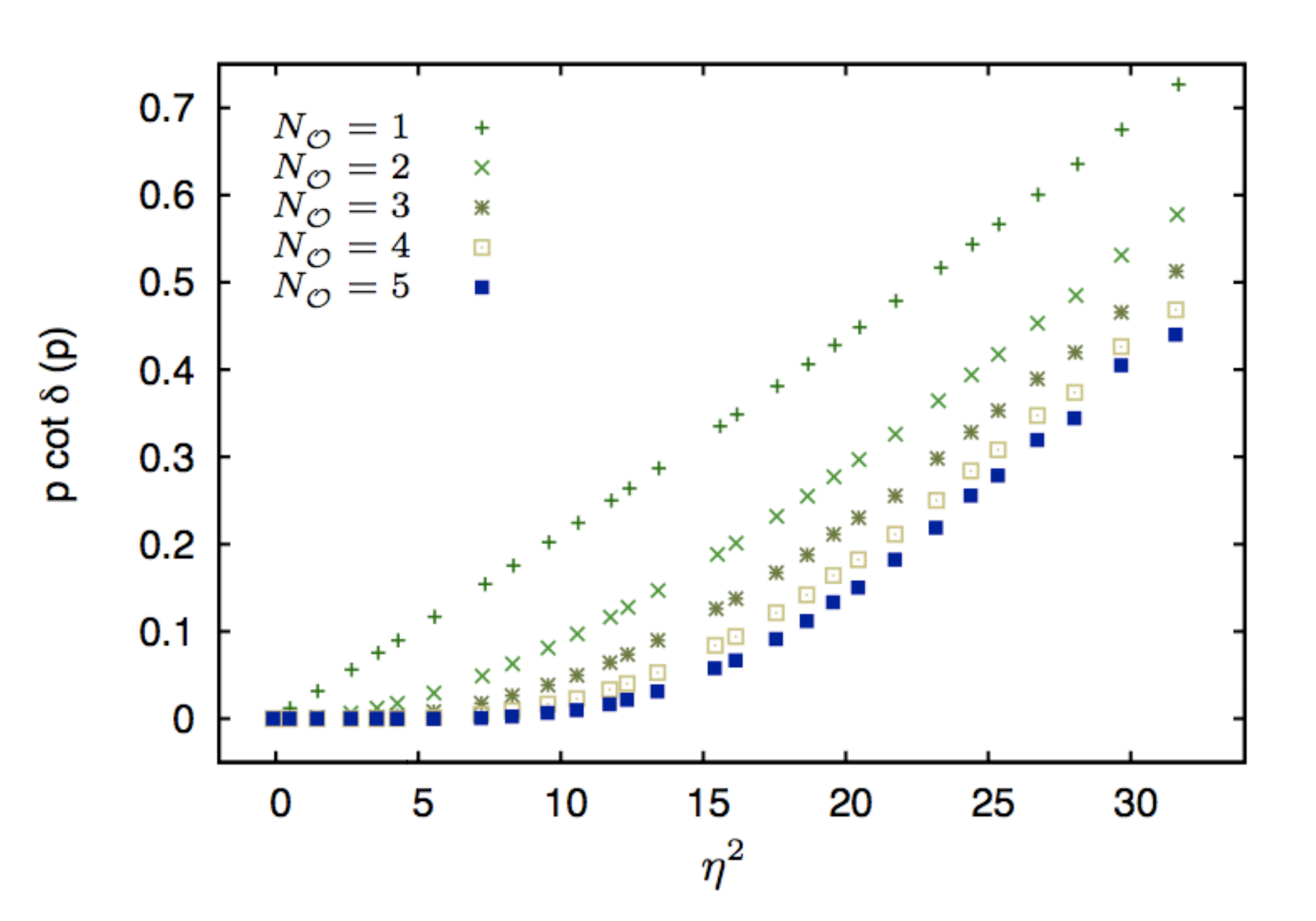}
\caption{\label{Fig:pcotdGI}(Color online) Plot of $p\cot \delta (p)$, in lattice units, as a function of 
$\eta^2 = {E^{}_2/p_0^2}$ (where $p_0^{} \equiv 2\pi/L$), for $N_x = 20$, $\tau = 0.05$, 
and levels of improvement $N^{}_\mathcal{O} = 1 - 5$, for a Galilean invariant definition of the transfer matrix.}
\end{figure}

Apart from the obvious desirability of Galilean invariance, these observations seem to indicate that one should choose 
this form over the non-invariant version. However, when comparing the approach to the unitary point based on improvements, 
as shown in Figs.~\ref{Fig:pcotd} and~\ref{Fig:pcotdGI}, it is clear that the non-invariant version is much 
superior to its invariant counterpart. Furthermore, comparing with the results of Ref.~\cite{Endresetal}, it seems clear that
one would need much larger lattices in the Galilean invariant version to make Fig.~\ref{Fig:pcotdGI} look like 
Fig.~\ref{Fig:pcotd}, which we have checked with our codes. (Note that Fig. 2 in Ref.~\cite{Endresetal}
presents the same kind of plot but corresponding to a much larger lattice size, namely $N_x = 32$.)

The reason for this difference is likely due to the fact that (discrete) Galilean invariance is only respected 
in an infinite lattice volume. Indeed, in any {\it finite} lattice the center-of-mass and 
relative motions are actually {\it not} separable. Boosting to frames of different total 
momentum will, in general, result in Hilbert spaces of vastly different dimensions for the subspace of relative motion. 
In particular, the zero total momentum Hilbert space, used to tune our two-body interaction, has by far the highest 
dimension. Based on this assessment, we decided to focus on the non-invariant form in this work.

Should one choose to implement the Galilean invariant form instead, as in Ref.~\cite{Endresetal}, 
there are two technical details that should not be overlooked. 
First, in Eq.~(\ref{2particleTmatrix}) the function $A({\bm p}_\uparrow^{})$ becomes 
$A({\bm p}_\uparrow^{} \!-\! {\bm q}_\uparrow^{})$ and therefore needs to be evaluated outside the 
single-particle momentum lattice where we originally defined it (c.f. Eq.~(\ref{Aexpansion})). 
The extension to the larger domain should be done respecting the periodicity of the momentum lattice. 

Second, if such a periodicity is to be reconciled with the continuum form of the 
non-interacting dispersion relation, which we assumed to be $E = p^2/2m$, we should impose a 
spherically symmetric cutoff $\Lambda$ in momentum space. 
This last condition should have little impact on the results in practice, as
long as the systems under consideration are somewhat dilute. From the point of view of 
computational performance, however, such a restriction on phase space results in important gains, 
particularly at finite temperature, where all the elements in the basis are evolved in imaginary time. 

Regardless of the form of the action, it is possible to evaluate the degree to which Galilean invariance is broken by
taking the improved transfer matrix and computing its eigenvalues in a moving frame. As long as 
Galilean invariance is respected, inserting the eigenvalues into the Rummukainen-Gottlieb formula~\cite{RummuGott}
should yield the same scattering phase shift as L\"uscher's formula. This will of course include breaking effects
coming from both the action {\it and} the lattice itself.

\begin{figure}[t]
\includegraphics[width=1.03\columnwidth]{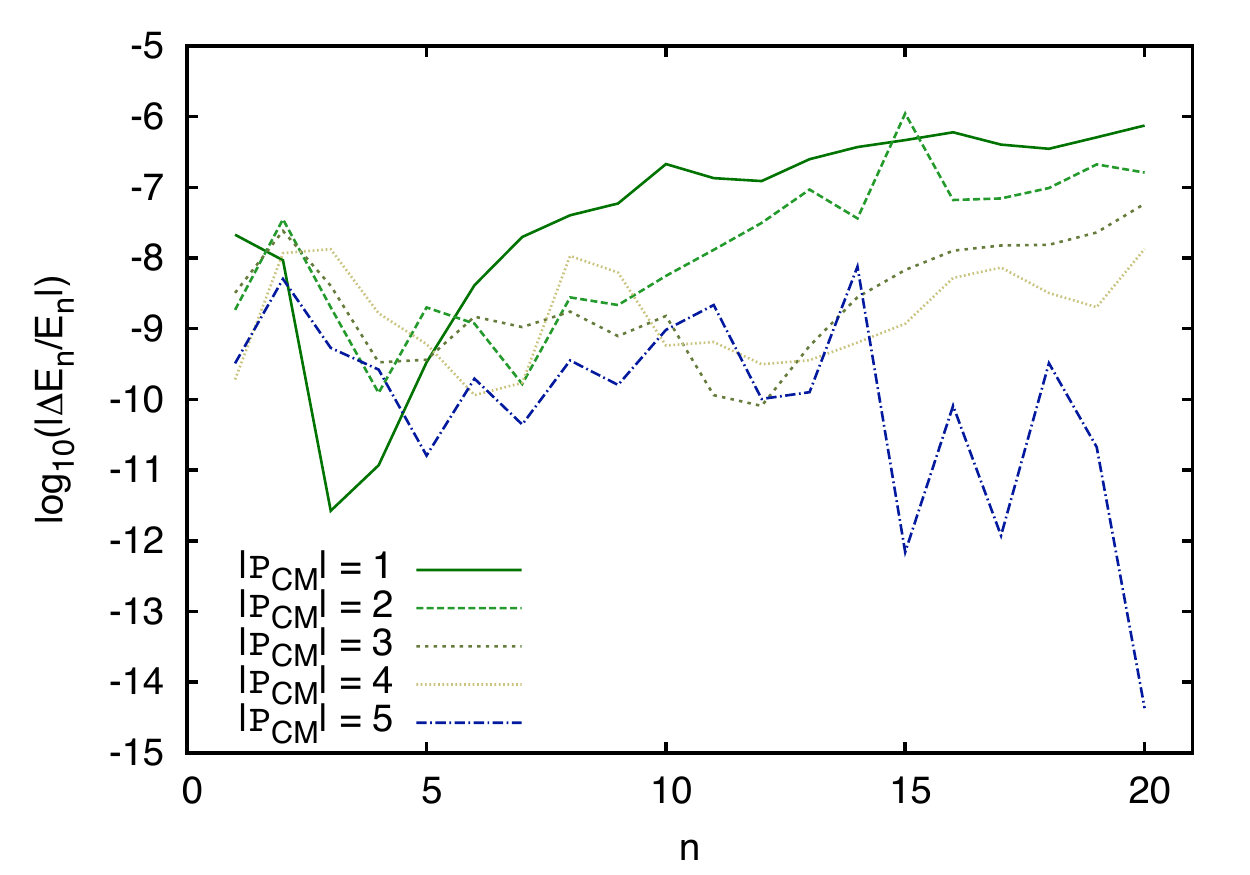}
\caption{\label{Fig:MovingFrameEigenvalues}(Color online) Logarithmic plot of the relative 
eigenvalue difference $|\Delta E^{}_n / E^{}_n|$ between the Galilean invariant and non-invariant 
improved transfer matrices, as a function of the eigenvalue index $n$, in various moving frames.
The quantity ${\bf P}^{}_\text{CM}$ denotes the center-of-mass momentum, in units of $2 \pi/L$.
This data set corresponds to $N_x = 18$, $\tau = 0.05$, $N^{}_\mathcal{O} = 3$.}
\end{figure}

Here, we limit ourselves to comparing the eigenvalues obtained from the invariant and non-invariant formulations,
in various frames. In Fig.~\ref{Fig:MovingFrameEigenvalues} we show a plot of the relative 
eigenvalue difference $|\Delta E^{}_n / E^{}_n|$ between the Galilean invariant and non-invariant 
improved transfer matrices, evaluated in various moving frames (i.e. non-zero center-of-mass momentum). 
As can be appreciated in that plot, the difference between the invariant and non-invariant improvements is 
extremely small, which shows that one can use the non-invariant version of the improvement program without 
introducing a noticeable systematic effect.



\end{document}